\documentclass[a4paper]{article}
\usepackage{icrc2013}
\title{The IFAE/UAB and LUPM Raman LIDARs for the CTA Observatory}
\shorttitle{IFAE/UAB Raman LIDARs}
\authors{A.~L\'opez Oramas$^1$, O.~Abril $^1$, O.~Blanch-Bigas$^1$, J.~Boix$^1$, V.~Da Deppo$^2$, M.~Doro$^{3,4,5}$, L.~Font$^{3,4}$, D.~Garrido$^{3,4}$, M.~Gaug$^{3,4}$, M.~Martinez$^1$ and G.~Vasileiadis$^6$ for the CTA consortium} 
\afiliations{
$^1$Institut de F\'isica d'Altes Energ\`ies (IFAE), Barcelona, Spain\\
$^2$CNR-Institute for Photonics and Nanotechnologies UOS Padova LUXOR, Padova, Italy\\
$^3$F\'isica de les Radiacions, Departament de F\'isica, Universitat Aut\`onoma Barcelona (UAB), 08193 Bellaterra, Spain \\ 
$^{4}$CERES, Universitat Aut\`onoma Barcelona-IEEC, 08193 Bellaterra, Spain\\
$^{5}$University and INFN Padova, via Marzolo 8, 35131, Padova, (Italy)\\
$^{6}$Laboratoire Univers et Particules de Montpellier (LUPM), Montpellier, France \\
}
\email{alopez@ifae.es}
\abstract{The Cherenkov Telescope Array (CTA) is the next generation of Imaging Atmospheric Cherenkov Telescopes. It will reach a sensitivity and an energy resolution with no precedent in very high energy gamma-ray astronomy. In order to achieve this goal, the systematic uncertainties derived from the atmospheric conditions shall be reduced to a minimum. Different instruments may help to understand and account for the precise state of the atmosphere at a given time. The Barcelona IFAE/UAB (acronyms for Institut de F\'isica d'Altes Energies and Universitat Aut\`onoma de Barcelona, respectively) and the Montpellier LUPM (Laboratoire Univers et Particules de Montpellier) groups are building Raman LIDARs, devices which can reduce the systematic uncertainties in the reconstruction of the gamma-ray energies from 20$\%$ down to better than 5$\%$. The Raman LIDARs will have each 1.8 m mirror and use a coaxial two-color beam from a Nd-YAG laser. A light guide collects the light at the focal plane and transports it to the readout system. We have developed a monochromatic optical detector with the purpose of testing the readout chain of both LIDARs. This device is composed of a system of filters and a photomultiplier, and will be used to study an individual elastic channel. After characterizing the system, we will build a multiwavelength optical detector to collect also the sparse Raman signal and will optimize it to reduce every possible loss of signal. We report on the current status of the LIDAR development and the latest results on the different characterization tests.}
\begin{document}
\maketitle
%Begin a section.
\section{Introduction}
Imaging Atmospheric Cherenkov Telescopes (IACTs) are ground-based telescopes which observe gamma rays in the GeV-TeV energy band. They detect the Cherenkov light produced by electromagnetic showers generated when primary gamma rays enter the atmosphere.\\

Cherenkov astronomy uses the atmosphere as a calorimeter, hence the knowledge of the atmospheric parameters is crucial \cite{bib:doro} . It is important to calculate the total extinction that the Cherenkov radiation suffers from the point where the primary gamma-ray interacts at 10-20 km height until it reaches the detector, in order to properly estimate the total flux and energy scale. At present, the systematic uncertainties in the determination of the energy scale at a given time for the current generation of IACTs is quoted at $\sim$20~\%.\\

The Cherenkov Telescope Array (CTA) \cite{bib:ctaconcept} will be the next generation gamma-ray observatory. It intends to improve the current sensitivity of IACTs by an order of magnitude and enlarge the energy range for the detection of gamma rays, covering almost four decades in energy. Additionally, it aims at considerably improved energy and angular resolution. For that purpose, it is necessary to reduce the systematic uncertainties as much as possible.\\

A way to determine the atmospheric parameters and perform a proper characterization of the atmosphere is by the use of a Raman LIDAR (which stands for \emph{Light Detection And Ranging}) \cite{bib:Weitkamp}, a device designed to monitor the atmospheric transmission with high precision \cite{bib:doro}, \cite{bib:garrido}.

\section{Raman LIDARs}
A LIDAR is composed of a powerful laser which points to the atmosphere, a mirror that collects the backscattered light and an optical detector. A Raman LIDAR collects not only the elastic Rayleigh scattered light but also the Raman component, produced by inelastic scattering on the air molecules. The study of both Rayleigh and Raman backscattered light will improve the knowledge of the height- and wavelength-resolved atmospheric extinction to better than 5~\%.\\

The amount of light scattered by the particles in the atmosphere will depend mainly on two factors: the attenuation suffered from the emission to the reception point and the backscatter cross section, which depends, among others, on the composition and shape of the aerosols. The backscattered signal is described by the LIDAR equation \cite{bib:monitoring}:\\

\begin{equation}
P(r,\lambda) = P_{0} {ct_{0} \over 2} \beta(r,\lambda) {A \over r^{2}} e^{-2 \tau(r,\lambda)}
\end{equation}

where P(r,$\lambda$) is the radiation received by the telescope, $P_{0}$ is the initial radiation emitted by the laser, c is the speed of light, $t_{0}$ is the time of the transmitted pulse, $\beta$(r,$\lambda$) is backscatter coefficient, where $r$ is the distance to the sensed part of the atmosphere, $A$ is the telescope mirror area and $\lambda$ is the wavelength. $\tau$(r,$\lambda$) represents the optical depth, which can be written in terms of the extinction, $\alpha$:\\

\begin{equation}
\tau(r,\lambda) = \int^{r}_{r_{0}}{\alpha(r,\lambda)} {dr}
\end{equation}

Eq. 1 contains two unknown parameters, namely $\beta(r,\lambda)$ and $\tau(r,\lambda)$, which cannot be retrieved simultaneously without further a priori assumptions 
on the relation between them. Furthermore, these unknowns will depend not only on the distance but also on the wavelength. The use of a Raman LIDAR will permit to break the degeneracy between $\beta$(r,$\lambda$) and $\tau$(r,$\lambda$) and, as a consequence, reduce the systematics due to the assumptions made otherwise. In the case of elastic LIDARs the backscatter power return depends on these two unknown parameters which need to be inferred from a single measurement. In this case, different boundary calibrations shall be introduced a priori, limiting the precision of the height-dependent atmospheric extinction to always worse than 20 ~\%. The introduction of additional elastic and/or Raman channels allow for simultaneous and independent measurement of the extinction and backscatter coefficients with no need for a priori assumptions \cite{bib:ansmann}. \\

%There is one equation with two unknown factors, $\beta$(r,$\lambda$) and $\tau$(r,$\lambda$), so approximations should be done. Furthermore, these unknowns will depend not only on the distance but also on the wavelength. The use of a Raman LIDAR will permit to break the degeneracy between $\beta$(r,$\lambda$) and $\tau$(r,$\lambda$) and, as a consequence, reduce the systematics due to the wavelength dependence. Note that, in the case of an 

The use of a Raman LIDAR permits, on one hand, the knowledge of the vertical distribution of aerosols, which would allow the decrease of the systematic errors derived from atmospheric quality knowledge. On the other hand, the monitoring of the atmospheric transmission at any moment would allow CTA to correct the data under non-optimal atmospheric conditions and therefore, increase the duty cycle.\\

%The IFAE/UAB and LUPM Raman LIDARs \cite{bib:icrc2011} are two similar LIDARs which are being developed at Barcelona and Montpellier, respectively.  Both contain a recycled CLUE container from the CLUE (\emph{Cherenkov Light Ultraviolet Experiment}) experiment, hosting a 1.8m diameter telescope. Both projects are following approximately the same approach for the LIDAR development. These LIDARs fullfil the requirements needed to be installed at the CTA observatory.\\

The Raman LIDARs subjects of this work will provide good angular resolution to determine extinction in terms of the height and the wavelength. They are sensitive to backscattered light from both elastic and Raman lines from a sensed region of the atmosphere up to 15 km height. In addition, a short integration time (of about 2-3 minutes) is required. For that purpose, a low signal-to-noise ratio is needed. These constrains lead to the need of using poweful lasers and large collection areas. As commercial LIDARs do not fullfil such requirements, dedicated LIDARs for CTA observatory have to be developed.

%\section{The IFAE/UAB and LUPM Raman LIDARs}

\section{Telescope structure}

%\subsection{Common Components}

The Raman lines are characterized by having a small cross section, about 2-3 order of magnitude smaller than the elastic cross section. Hence, a big reflecting area is needed to collect a substantial amount of light. Studies of the PSF and reflectivity show that the mirrors have maintained a good optical quality, keeping basically the same PSF as when they were produced ( $\approx$ 6.5 mm) for CLUE (\emph{Cherenkov Light Ultraviolet Experiment}, former experiment placed at Roque de Los Muchachos observatory in La Palma.), although the reflectivity has decreased to 64 ~\% at 350 nm.\\

\begin{figure}[th!]
\centering
\includegraphics[width=0.5\textwidth]{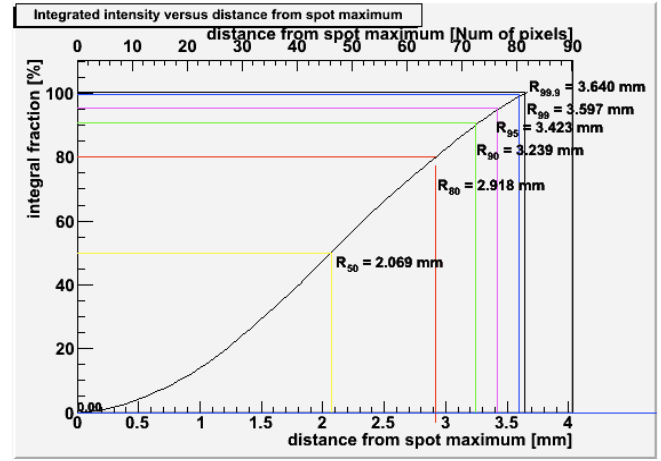}
\caption{ Measurement of the primary mirror PSF by using an artificial star. All light is enclosed in the liquid light guide perfectly. In this case, D{$_{90}$} $\sim$ 6.4 mm}
\label{icrc2013-0210-01}
\end{figure}

The mirror spot size has been characterized through its PSF by different methods: pointing to Polaris and using an artificial star created with a laser pointing. Different energy containments at different radii are calculated.The results obtained with the artificial star show that 80$\%$ of the light is contained in an area of $\sim$6mm diameter, while the 90$\%$ falls in a circle with a diameter of about 6.4 mm (see Fig.~\ref{icrc2013-0210-01} ). \\

Both telescopes have been equipped with Nd:YAG lasers, which can provide three different wavelengths, 1064nm, 532nm and 355nm. The three wavelengths will be emitted simultaneously. The lasers are mounted on a two-dimensional XY table which corrects the position by feedback.\\

\begin{figure}[t]
\centering
\includegraphics[width=0.45\textwidth]{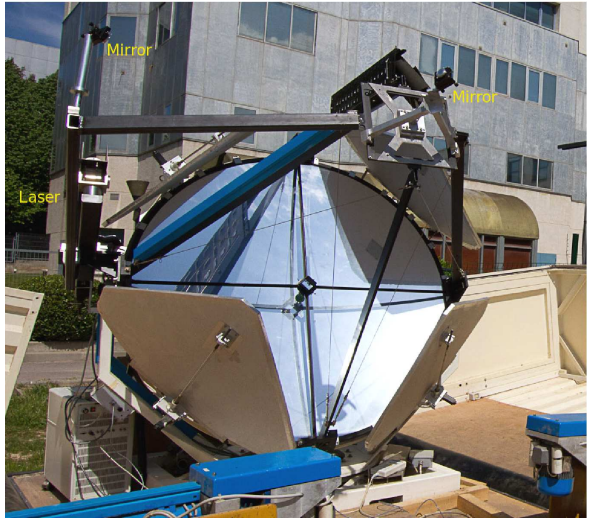}
\caption{Telescope used for the IFAE/UAB LIDAR development. The supporting arm for the laser is visible on the left side. Two flat mirrors will guide the light emitted by the laser in such a way that it will be coaxial with respect to the main axis of the primary mirror. The LLG is on the focal plane (before the mirror) and will transport the light until the back part of the structure, where the optical system will be placed.}
\label{telescope}
\end{figure}

%\subsection{Coaxial Setup}
A coaxial configuration has been set for the IFAE/UAB and LUPM Raman LIDARs (see  Fig.~\ref{telescope}) in order to achieve a good overlap between the laser beam ligth cone and the telescope’s field of view. A good overlapping factor is needed to measure the optical depth starting from a few tens of meters height. It is important  to start measuring from low altitudes to perform a proper atmospheric characterization, as there is where the atmosphere is more hazy. We will describe the coaxial configurations set for both IFAE/UAB and LUPM LIDARs.  \\

As the IFAE/UAB laser cannot be placed in the center of the colleting area, it relies on a mechanical arm which guides the laser beam to the desired direction and position. Two flat highly resistant 1-inch mirrors have been placed, one in front of the laser (at a distance of about 1.5 m) and the other in the focal plane, in order to guide the laser light in such a way that it becomes coaxial with the optical axis of the primary mirror. These two mirrors are strongly and precisely fixed on the telescope structure, on the positions of maximum alignment, where the coaxial configuration and hence a good overlapping factor are achieved,. These guiding mirrors can be displaced from this initial position and can be replaced along the structure in case that a new configuration is to be set. The movements of the LIDAR will cause a certain misalignment of the system. For that reason, the mechanical arm supporting the laser has been provided with two stepping motors which can correct these deviations. First, a pre-alignment is performed which lasts about 10 min. Then, the laser return signal at large distances is maximized by moving the laser beam.\\

%At first, it makes an initial alignment proceeding which will last about 10 min, then the laser shoots to four points and, depending of the return, it moves to the position which gave it the best value until it reaches the best approach.\\

Liquid light guides (LLG) are used in both IFAE/UAB and LUPM Raman LIDARs to transport the light from the focal plane to the detector, due to their high transmitivity and easy handling. A support for the light guide has also been installed in the structure of the telescope. The pieces fix it to the structure in such a way that movements are restricted and a minimum bending radii for the LLG is assured.\\

%can move and adapt to the movements of the telescope assuring a minimum bending radii.\\

The approach for the LUPM LIDAR is slightly different. In this case, the laser head is firmly mounted on the telescope dish. A single 1-inch fixed mirror will guide the laser beam to the center of the dish at the level of the focal point. An XY software controllable table will be used at this point to align the laser beam with the optical axis of the telescope.\\

%The coaxial configuration set for teh IFAE/UAB LIDAR is shown in Fig.~\ref{telescope}

\section{Optical detector}
Currently in its construction phase, the optical detector is a device which will collect the light from the output of the LLG, split the beam into the different wavelengths of interest and focus each of them onto a photomultiplier (PMT). The design has been done in such a way that the dimension and the number of optical components was the smallest possible and the efficiency was maximum \cite{bib:polychromator}.\\

\subsection{Monochromatic optical detector}

%It is worth to understand the behavior of our system before using a complex device as multiwavelength optical detector is. 

\begin{figure}[th!]
\centering
\includegraphics[width=0.5\textwidth]{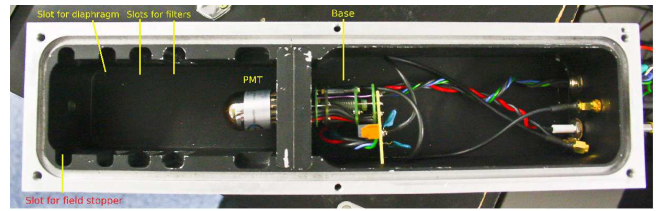}
\caption{Monochromatic optical detector.}
\label{monochromator}
\end{figure}

 A single channel detector has been built for the IFAE/UAB Raman LIDAR , the monochromatic detector, which has been developed in order to test the LIDAR response. It has been built only for testing purposes, then a multiwavelength detector will be used. It is a simple instrument which hosts a photomultiplier, a filter (although currently there are two slots for optical elements ) and a diaphragm (Fig.~\ref{monochromator}). The backscatter cross section for the Raman lines is 2-4 orders of magnitude smaller than the Rayleigh one. Because the intensity of the Rayleigh scattered light is not a limitation, it will collect the light from a single elastic channel.

%It collects the light from a single elastic channel, because there we are not limited by the amount of light. It is worth to remember that the backscatter cross section for the Raman lines is by 2-4 orders of magnitude smaller than the Rayleigh one.\\

The readout chain, the optics and all elements will be tested with this device. It will help to understand the full performance of the LIDAR and will allow the optimization of the system. It will first be tested in the laboratory to understand the behavior in a controlled environment. Then, it will be mounted on the LIDAR and will be operated to test the real response of the system during night operation.\\

%This detector will be used to test the full performance of the system: optics, readout chain... It has been first tested and characterized in the laboratory and then it has worked with the LIDAR at night. It has been developed in order to test the LIDAR response. It will help to understand the performance of our device and will permit the optimization of the system.\\

%that is the purpose of the monochromator. First it has been tried in the laboratory and once characterized, it has been tested with the LIDAR during night operation. In order to test the readout chain of the IFAE/UAB Raman LIDAR, a single channel detector called \emph{monochromator} has been built.\\

%The monochromatic detector collects the light from a single elastic channel, because there we are not limited by the amount of light. It is worth to remember that the backscatter cross section for the Raman lines is by four orders of magnitude smaller than the Rayleigh one.\\

%As a test bench, a former MAGIC PMT with single photo-electron response has been used. Both the AC and DC coupling signal can be read with the current electronics. The PMT is fixed tightly inside a metallic box. 

\subsection{Multiwavelength optical detector}
A four-channel optical detector composed of lenses, dichroic mirrors, interference filters and PMTs has been designed \cite{bib:polychromator}. There will be two channels for the elastic lines at 355 nm and 532 nm and another two channels for the Raman lines of the nitrogen at 387 nm and 607 nm. As mentioned before, it is important to have a system with at least one Raman channel, to break the degeneracy in the LIDAR equation and retrieve the atmospheric extinction to a precision better than 5~\%. The introduction of additional channels will help in the reconstruction of the optical properties of the atmosphere. Signals measured at different wavelengths can be used to optimize the LIDAR equation solution.  A second elastic channel helps to determine the LIDAR calibration parameters and decreases the uncertainties in the atmospheric transmission calculation from 20~\% to roughly 15~\%. Furthermore, a second Raman channel  permits to separate the effect of molecules and aerosols in the backscattered coefficient, which affects the total return power.\\

\begin{figure*}[!ht]
\centering
\includegraphics[width=0.8\textwidth]{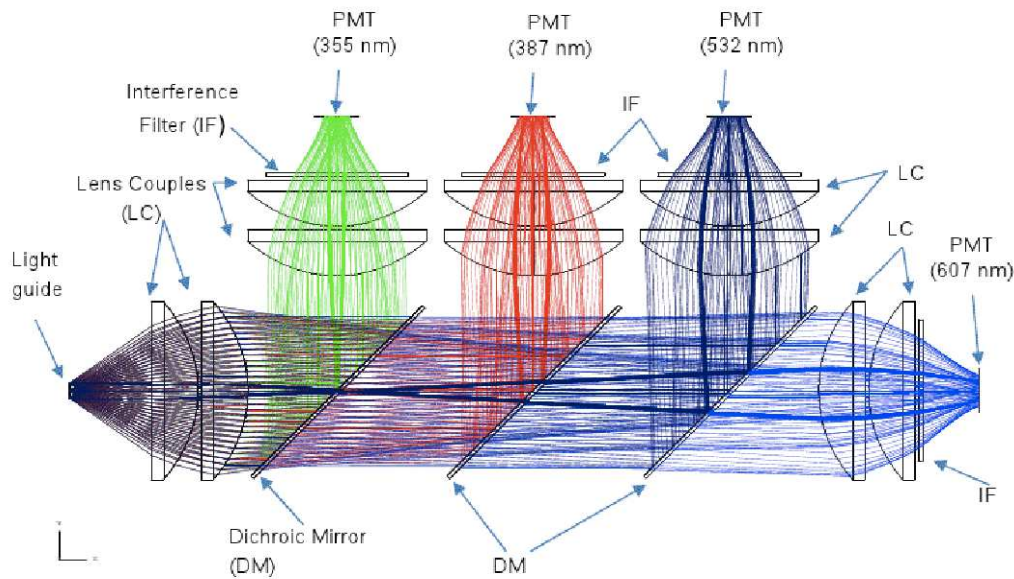}
\caption{Four-channel optical detector designed for the Raman LIDARs of this paper. The light coming from the light guide is well collimated onto each PMT.}
\label{icrc2013-0210-04}
\end{figure*}

%be able to make a proper estimation of the extinction.\\
The final design for the optical system for the multiwavelength detector is shown on Fig.~\ref{icrc2013-0210-04}. The light which arises from the light guide at the detector entrance has a wide opening angle (70$^o$ aperture angle) due to the f/D = 1 relation of the primary mirror. Hence, collimation and focalization of the beam onto the PMT is not straightforward. Due to this reason, a couple of lenses are placed to collimate the beam. Then, dichroic mirrors will separate the light into the wavelengths of interest. Now that the channels are differentiated, each single beam will go through another pair of lenses which will help to collimate it into the PMT and through an interference filter to improve the wavelength selection. Finally, the single-wavelength light arrives at the PMT. For each channel, 80~\% of the energy is enclosed inside the effective area of the PMT. \\

%The final design for the multiwavelength detector is shown in Fig.~\ref{icrc2013-0210-04}

\section{Readout}

Each channel shall read a return power which spans for several orders of magnitude. There will be a continuous signal recording for the lowest altitudes and single-photon detection for the highest (around 15 km) ones. The data acquisition provides DC readout for the low regime and single photo-electron counting for the high altitudes. Therefore, also the PMTs must provide single photo-electron solution. Both Raman LIDARs are using commercial standard LICEL modules for the readout, although the idea of a homemade solution is being discussed.

\section{Outlook and conclusions}

The future Cherenkov Telescope Array will be the next generation gamma-ray observatory. It will have an order of magnitude greater sensitivity relative to current IACTs and a considerably larger sensitive energy range. For that purpose, a proper characterization of the atmosphere has to be done, in order to estimate the extinction that the Cherenkov photons undergo between the shower emission height and the ground and correctly reconstruct the energy of the primary gamma ray.
%in order to estimate the extinction that the UV photons undergo from the top of the atmosphere to the ground and correctly reconstruct the energy of the primary gamma ray.\\

A suitable device to achieve this goal is a Raman LIDAR, an instrument that can provide a height- and wavelength-resolved characterization of the atmosphere. The Raman LIDAR would allow to correct systematic biases on the energy scale and flux, reducing the systematic uncertainties from 20~\% down to better than 5~\%. This will help not only to reach the desired energy resolution but also to increase the duty cycle, because the observatory will be able to work under non-optimal weather conditions.\\

The IFAE/UAB and LUPM Raman LIDARs are being designed in Barcelona and Montpellier for that purpose. Both systems have finished their designed phase and they are starting to become operational. The optical system is currently under construction and a first test bench is being used both in the laboratory and with the real system.  As soon as the construction of the optical system is complete and its behavior is understood, the Raman LIDARs will be fully operational and ready for deployment to the CTA site.

%As soon as the behavior is completely understood and the multiwavelength optical constructed, the LIDARs will be operational as full Raman LIDARs.

\section{Acknowledgments}

We gratefully acknowledge support from the following agencies and organizations:
Ministerio de Ciencia, Tecnolog\'ia e Innovaci\'on Productiva (MinCyT),
Comisi\'on Nacional de Energ\'ia At\'omica (CNEA) and Consejo Nacional  de
Investigaciones Cient\'ificas y T\'ecnicas (CONICET) Argentina; State Committee
of Science of Armenia; Ministry for Research, CNRS-INSU and CNRS-IN2P3,
Irfu-CEA, ANR, France; Max Planck Society, BMBF, DESY, Helmholtz Association,
Germany; MIUR, Italy; Netherlands Research School for Astronomy (NOVA),
Netherlands Organization for Scientific Research (NWO); Ministry of Science and
Higher Education and the National Centre for Research and Development, Poland;
MICINN support through the National R+D+I, CDTI funding plans and the CPAN and
MultiDark Consolider-Ingenio 2010 programme, Spain; Swedish Research Council,
Royal Swedish Academy of Sciences financed, Sweden; Swiss National Science
Foundation (SNSF), Switzerland; Leverhulme Trust, Royal Society, Science and
Technologies Facilities Council, Durham University, UK; National Science
Foundation, Department of Energy, Argonne National Laboratory, University of
California, University of Chicago, Iowa State University, Institute for Nuclear
and Particle Astrophysics (INPAC-MRPI program), Washington University McDonnell
Center for the Space Sciences, USA. The research leading to these results has
received funding from the European Union's Seventh Framework Programme
([FP7/2007-2013] [FP7/2007-2011]) under grant agreement nÂ° 262053.\\

%We gratefully acknowledge financial support from the agencies and 
organizations listed in this page: http://www.cta-observatory.org/?q=node/22\


\begin{thebibliography}{}

\bibitem{bib:doro} M. Doro, M. Gaug, O. Blanch, Ll. Font, D. Garrido, A. L\'opez-Oramas, M. Mart\'inez, \emph{Towards a full Atmospheric Calibration system for the Cherenkov Telescope Array}, these proceedings, ID 0151.

\bibitem{bib:ctaconcept} Acharya, B.S. et~al., Astrop. Phys., 43:3--18, 2013.

\bibitem{bib:Weitkamp} Weitkamp, C. editor, 2005, \emph{LIDAR Range-Resolved Optical Remote Sensing of the Atmosphere}, Weitkamp, C., Springer 

\bibitem{bib:garrido} D.,Garrido, et al., \emph{Influence of atmospheric aerosols on the performance of the MAGIC telescopes}, these proceedings, ID 0465

\bibitem{bib:monitoring}
Mussa R. et al, Nuclear Physics B, 2009, Volume 190, 272-277

\bibitem{bib:ansmann} A. Ansmann, U. Wandinger, M. Riebesell, C. Weitkamp, and W. Michaelis. \emph{Independent measurement of extinction and backscatter profiles in cirrus clouds by using a combined raman elastic-backscatter lidar}. Appl. Opt., 31:7113, 1992.

\bibitem{bib:icrc2011} M. Barcel\'o et al, \emph{Development of Raman LIDARs made with former CLUE telescopes for CTA}, DOU: 1-.7529/ICRC2011/V09/0408

\bibitem{bib:polychromator} V. Da Deppo et al, Proc. of SPIE, Vol. 8550, 85501V


\end{thebibliography}
\end{document}